\documentclass[final,5p,times,twocolumn,authoryear]{elsarticle}
\pdfoutput=1

\usepackage{hyperref}
\hypersetup{
    colorlinks=true,
    linkcolor=myblue,
    citecolor=myblue,
    urlcolor=myblue
}

\usepackage[pagewise]{lineno}
\usepackage{newtxtext,newtxmath}
\usepackage{subcaption}
\usepackage{xcolor}
\usepackage{multirow}
\usepackage{threeparttable}
\usepackage{subcaption}
\usepackage{mwe}
\usepackage{comment}
\usepackage{adjustbox}
\usepackage{graphicx}
\usepackage{amsmath}
\usepackage{footnote}
\makesavenoteenv{table}
\makesavenoteenv{tabular}
\usepackage{tabularx}
\usepackage{graphicx}
\usepackage{soul}
\usepackage{epstopdf}

\usepackage[utf8]{inputenc}
\usepackage{textcomp}
\usepackage{newunicodechar}
\newunicodechar{−}{\textminus}

\usepackage{float}
\usepackage{lipsum}
\usepackage{pdfpages}

\usepackage[utf8]{inputenc} 
\usepackage[T1]{fontenc} 

\usepackage{titlesec}
\titleformat{\subsection}
  {\normalfont\bfseries}{\thesubsection}{1em}{}

\newcommand{\src}{PSR J1744--1134 }

\begin{document}
\date{Accepted XXX. Received YYY; in original form ZZZ}
{
\makeatletter
\renewcommand*{\corref}[1]{\@gobble}
\providecommand*{\cnotenum}[1]{}
\makeatother

\begin{frontmatter}
\journal{Journal of High Energy Astrophysics}
\title{Study of pulse jitter characteristics of PSR J1744$-$1134$:$ Towards high precision pulsar timing with InPTA}

\author[label1]{Jibin Jose}
\ead{phd2301121007@iiti.ac.in}

\author[label1]{Manoneeta Chakraborty}


\author[label2]{Neelam Dhanda Batra}

\author[label3]{K Nobleson}


\author[label4]{Shantanu Desai}

\author[label5,label6]{Manjari Bagchi}

\author[label7,label8]{Bhal Chandra Joshi}

\author[label3]{Keitaro Takahashi}

\author[label9]{Shaswata Chowdhury}

\author[label10]{Churchil Dwivedi}

\author[label11]{Abhimanyu Susobhanan}

\author[label12]{Debabrata Deb}

\author[label13]{Sushovan Mondal}

\author[label14]{Zenia Zuraiq }

\author[label15]{Kaustubh Rai}

\author[label6]{Sharika Dhakappa}

\author[label16]{Pratik Tarafdar}

\author[label17]{Kuldeep Meena}

\address[label1]{Department of Astronomy, Astrophysics, and Space Engineering, Indian Institute of Technology Indore, Indore 453552, India}

\address[label2]{Department of Physics, IIT Delhi, Hauz Khas, New Delhi 110016}

\address[label3]{Faculty of Advanced Science and Technology, Kumamoto University, 2-39-1 Kurokami, Kumamoto 860-8555, Japan}



\address[label4]{Department of Physics, IIT Hyderabad, Kandi
Telangana-502284, India}

\address[label5]{The Institute of Mathematical Sciences, C. I. T. Campus, Taramani, Chennai 600113, India  }

\address[label6]{Homi Bhabha National Institute, Training School Complex, Anushakti Nagar, Mumbai 400094, India}

\address[label7]{National Centre for Radio Astrophysics, SP Pune University Campus, Pune, Maharashtra, 411007, India }

\address[label8]{Department of Physics, Indian Institute of Technology Roorkee, Roorkee, Uttarakhand, 247667, India }

\address[label9]{Radio Astronomy Centre,Tata Institute of Fundamental Research, National Centre for Radio Astrophysics, Ooty, Tamil Nadu 643001}

\address[label10]{Astronomy and Astrophysics Division, Physical Research Laboratory, Thaltej Campus, Thaltej, Ahmedabad, 380059, Gujarat, India}

\address[label11]{School of Physics, Indian Institute of Science Education and Research Thiruvananthapuram, Maruthamala PO, Thiruvananthapuram, 695551, Kerala, India}

\address[label12]{Centre for Space Research, North-West University, Private Bag X6001, Potchefstroom 2520, South Africa}

\address[label13]{Department of Physics, Indian Institute of Technology Bombay, Mumbai, 400076, India.}

\address[label14]{Department of Physics, Indian Institute of Science, Bengaluru, 560012, India}

\address[label15]{Indian Institute of Science Education and Research, Bhopal Madhya Pradesh, 462006,  India}

\address[label16]{INAF - Osservatorio Astronomico di Cagliari, via della Scienza 5, 09047 Selargius (CA), Italy}

\address[label17]{UM-DAE Centre for Excellence in Basic Sciences,  University of Mumbai, Vidyanagari, Mumbai 400098, India.}

\begin{abstract}
Variations in the shapes of individual pulses introduce a broadband noise component in pulsar times of arrival beyond what is expected from radiometer noise, referred to as the jitter noise. This noise limits the timing accuracy and, consequently, the sensitivity required for the detection of nanohertz gravitational waves using pulsar timing arrays. Accurate characterization of jitter noise, including its amplitude and frequency dependence, is therefore essential for each pulsar. In this work, we present the first detailed study of jitter noise in PSR J1744$-$1134 over the frequency range 300–500 MHz, using observations from the Indian Pulsar Timing Array experiment conducted with the upgraded Giant Metrewave Radio Telescope. We detect clear signatures of jitter in this pulsar through pulse-level analysis and Bayesian inference. Furthermore, we investigated the frequency dependence of the jitter noise and compared our results with earlier measurements at higher frequencies. The observed differences in jitter amplitude suggest a possible frequency evolution, highlighting the importance of low-frequency studies for improving pulsar timing precision and accuracy. We also investigated the effect of jitter on the high precision dispersion measure derived from our observations and found that jitter is not the limiting factor for DM precision achieved for this pulsar in the given observation timescale.
\end{abstract}


\end{frontmatter}
}

\section{Introduction}
\label{sec:intro}

The pulsar timing array (PTA) sensitivity to gravitational waves (GWs) depends on the various sources of noise present in the data. The sources of noise that influence the measured time of arrivals (TOAs) can be categorized based on several characteristics, such as time correlation and frequency dependence. Time correlation describes whether the arrival-time fluctuations persist over extended observation epochs or act independently from pulse to pulse. Long-term correlated processes, where power is concentrated at low fluctuation frequencies, are collectively known as red noise; these arise from rotational irregularities of the pulsar, temporal variations in the dispersion measure (DM), and errors in the solar system ephemeris. In contrast, noise with zero temporal memory between epochs is called white noise, primarily driven by thermal radiometer noise and single-pulse jitter noise. Noise processes are further characterized by their radio-frequency dependence, or chromaticity, which describes how the delay scales across different observing bands. Processes like Solar System ephemeris errors and intrinsic spin noise are achromatic , delaying all radio frequencies equally. Conversely, propagation effects through the ionized interstellar medium  such as dispersive delays  and plasma scattering  are strongly frequency-dependent. Finally, noise can be categorized by whether it is common to multiple pulsars or unique to an individual pulsar. Modeling these noise sources is critical for extracting the GW signal from PTA datasets~\citep{Verbiest2018}.

At the individual pulse level, each pulse differs in terms of amplitude, shape, and phase. These differences can introduce noise into the integrated pulse profile, which is obtained by averaging a large number of individual pulses. Its effect is visible in the timing residuals as stochastic fluctuations with a white noise spectrum, known as pulse jitter \citep{Oslowski11}, \citep{Parthasarathy2021}. On short time scales, the excess noise in the timing residuals (noise beyond timing model) is dominated by radiometer noise and jitter noise. Radiometer noise represents the fundamental thermal noise floor of the receiver system, arising from Gaussian voltage fluctuations generated by receiver electronics, atmospheric emission, and background sky temperature \citep{Wilson2013}. Its amplitude in flux density scales inversely with the square root of the system resources according to the radiometer equation, $\sigma_S =\frac{T_{\text{sys}}}{ G\sqrt{n_{\text{pol}}} \Delta \nu T}$, where $T_{\text{sys}}$ is the system temperature, $n_{\text{pol}}$ is the number of polarizations, $\Delta \nu$ is the bandwidth,$T$ is the integration time and G is the gain of telescope \citep{Lorimer2004}. The amplitude of the radiometer noise depends on the observing system configuration, and is always uncorrelated in time and homoscedastic across the pulse phase bins. On the contrary, jitter noise in different phase bins will be temporally correlated and heteroscedastic.
To achieve the science objectives of pulsar timing arrays, particularly the detection of nanohertz gravitational waves, it is essential to improve the sensitivity and precision of pulsar timing by minimizing the contributions from both radiometer and jitter noise. As both are white noise processes, their Root Mean Square (RMS) amplitudes scale as $T^{-0.5}$, where $T$ is the sub-integration time over which pulses are averaged to suppress radiometer noise and intrinsic pulse-to-pulse variability. By increasing the sub-integration time, the  amplitude of these noise sources  can be reduced but their relative dominance cannot be changed, because both noise sources have the same scaling property with sub-integration time. So for the case where jitter noise dominates,  a separate jitter noise modeling is needed. Otherwise, the uncertainty associated with each TOA will be underestimated, which will affect the subsequent noise analysis. So, once the observation strategy hits the jitter noise floor, there is no further benefit to increasing the sub-integration time to reduce these noise effects. This means jitter noise plays an important role in determining both the observation time and the sub-integration time adopted for a pulsar. Different PTAs adopt different sub-integration times, based on operational constraints, RFI (radio frequency interference) excision, maintaining phase stability, and yielding sufficient epochal TOAs for multichannel noise analysis.  


In the near future, with high-sensitivity telescopes such as the Square Kilometre Array (SKA), it will be possible to obtain sufficient signal-to-noise ratio (S/N) within a small observation time~\citep{SKAO25}. 
The enhanced sensitivity of SKA facilitates the characterization of single-pulse variability, resulting in subtle differences between the averaged pulse profile and the template, the effects of which appear as jitter noise in timing analyses. 
As next-generation PTA experiments plan to adopt shorter observation durations owing to the higher sensitivity of the radio telescopes employed by these experiments, they may become increasingly affected by jitter noise, if its impact on individual pulsars is not well characterized. It is therefore important to investigate the properties and amplitude of the jitter noise for each observed pulsar to design optimal observing strategies.

This work is a follow-up to our previous study of jitter for PSR J0437$-$4715~\citep{Kikunaga2024} using InPTA observations. \citet{Kikunaga2024} reported a linear increase  in jitter with frequency within the frequency range 300$-$500 MHz. This behavior is opposite to the linear decreasing trend of jitter with frequency reported by \citet{Parthasarathy2021} in the frequency range 900$-$1600 MHz. This suggests the possible existence of a \textit{turnover} frequency region where the jitter trend reverses with frequency. The results of that study motivated us to investigate the jitter noise in other pulsars, characterize its frequency dependence, and the possible existence of such \textit{turnover} frequencies. In this work, we study the jitter noise in the pulsar PSR~J1744$-$1134. \par 
PSR J1744$-$1134, discovered in the Parkes 436 MHz southern sky survey,  has a spin period of 4.07 ms \citep{Bailes_1997}. This bright MSP exhibits a highly coherent pulse profile with a 10\% duty cycle. Hence, this is a source of interest to all PTAs, and   is monitored regularly. In particular, its steep spectral index and high modulation index (fluctuation of pulse intensity from one pulse to the next, normalized by the average intensity) make it an ideal candidate for timing and jitter studies at low frequencies~\citep{Gitika_2023}, and thus it is regularly observed as part of the  InPTA experiment. PSR~J1744$-$1134 has also been studied by multiple groups to characterize pulse jitter at L-band frequencies, with observations reported at 1284~MHz \citep{Parthasarathy2021}, 1500~MHz \citep{Lam2019}, 1400~MHz \citep{shannon_2014, Lam2016}, and 1250~MHz \citep{wang2024}, as part of large-scale jitter and pulsar timing studies. \citep{Lam2019} also reported a jitter amplitude measurement at 820~MHz. All of the previous L-band investigations were conducted at frequencies above 1~GHz, while the 820~MHz measurement provides an additional lower-frequency point. In contrast, this work presents the first detailed investigation of pulse jitter below 500~MHz.
In this paper, we investigate the jitter amplitude of PSR J1744$-$1134 in the frequency range 300$-$500 MHz using the upgraded Giant Metrewave Radio Telescope (uGMRT,~\citealt{Gupta2017}) data. The paper is structured as follows. In Section~\ref{sec:obs}, we discuss uGMRT observations, data processing methods, and methodologies used for epoch selection.  In Section~\ref{sec:analysis}, we discuss different methodologies for characterizing  jitter and obtain corresponding results. Finally, we present the physical implications of our work, the comparison with the jitter properties of pulsars reported in the literature, and the conclusions, in Section~\ref{sec:disc}.

\section{Observation, Data processing and Epoch selection}
\label{sec:obs}
\subsection{Observations}
We briefly summarize the observations and the associated data reduction techniques used for the analysis in this work. More details can be found in our previous works~\citep{Tarafdar22,Rana25}.

The InPTA observations are carried out with the uGMRT in total-intensity mode, where the GMRT Wideband Backend (GWB) \citep{Reddy} records frequency-resolved time-series data. 
They were carried out during observation Cycles 42 to 47 of uGMRT, where the source is observed either using two sub-arrays of 10 to 15 antennae at Band 3 (300-500 MHz) and Band 5 (1260$-$1460 MHz) or using 20 antennae at Band 3 in phased array mode. The Band 3 data (300$-$500 MHz) were coherently de-dispersed using the known dispersion measure ($DM = 3.1387$ pc $cm^{-3}$) of the pulsar to 128 frequency channels using a real-time pipeline and sampled every $5 \mu s$~\citep{De2016}. The Band 5 data were recorded in the regular phased array mode and sampled at 40.96 $ \mu s$. Both Band 3 and Band 5 data are stored after averaging to 10 seconds.

\subsection{Data processing}
This subsection outlines the data processing procedures adopted in the InPTA experiment and the software tools used for data reduction. The raw observational data were  processed using the \texttt{PINTA}~\citep{pinta_2021}
pipeline which performs RFI mitigation and partial folding using \texttt{RFIClean}~\citep{RFIClean_2021} and \texttt{DSPSR}~\citep{vanStraten2011} respectively. This produces \texttt{PSRFITS} archives \citep{Hotan2004}. Later,  \texttt{DMCalc}~\citep{Kk_2021} provides the  estimates of TOAs (Time of arrival) and DMs.

Since the jitter analysis assumes that deterministic timing analysis has accounted for all the known systematic effects, we updated the header of each PSRFITS file using the Data Release 2 (DR2~\citealt{Rana25}) timing solutions  and the DM values obtained for each epoch with \texttt{DMCalc}. The \texttt{PSRFITS} files were then de-dispersed and frequency-collapsed to eight sub-bands using the \texttt{pam} command in \texttt{PSRCHIVE} \citep{hotan_2004,van_2012}. The TOAs for each partially folded profile (averaged over 10 s of data) were determined by cross-correlating them with a frequency-resolved template using the \texttt{pat} command in \texttt{PSRCHIVE}. The same template employed in the DR2 timing analysis was also used for cross-correlation  \par
Using the \texttt{TEMPO2} software package~\citep{hobbs_2006,edwards_2006}, we obtained timing residuals for further analysis. This software compares the observed TOAs with the predicted TOAs from the model that contains rotational, astrometric, and binary parameters. The sum of squares of the differences between the observed and predicted TOAs, called timing residuals, is minimized to obtain the best fit parameters of the pulsar. Since we are using the timing solution from DR2, we are not doing any fitting of parameters. However, we removed all TOAs with uncertainties greater than 10 $\mu s$, as the improved TOAs now available render these lower-precision measurements unlikely to contribute useful information to the analysis. The \texttt{tim} file (file that contains TOAs of pulsar pulses) made after this is used for further analysis that is described in the next section.

\subsection{Epoch selection}
The S/N threshold was chosen to identify the regime in which the timing precision is no longer limited by radiometer noise. The TOA uncertainty, $(\sigma_{\rm TOA})$, which represents the radiometer-noise contribution, is inversely proportional to the S/N. Therefore, observations with higher S/N have relatively smaller contribution from radiometer-noise uncertainties. If the observed TOA scatter is comparable to $\sigma_{\rm TOA}$, the timing precision is radiometer-noise limited, making it difficult to reliably characterize any additional noise process. As the S/N increases, the radiometer-noise contribution decreases, allowing other noise sources, such as pulse jitter, to become measurable. To determine the S/N above which jitter can be reliably characterized, we examined whether the estimated jitter amplitude (on the assumption that epoch is jitter dominated) exhibited the expected behavior of pulse jitter, namely the $1/\sqrt{T}$ scaling with sub-integration time and approximate independence on observing bandwidth. We found that below a one-hour scaled S/N of 1139 (one hour scaling was performed for comparison as the observation time of each epoch was different; the one-hour scaled S/N comparison gives an unbiased selection of epochs), the estimated jitter amplitudes no longer followed these expected trends, indicating that the measurements were dominated by radiometer noise. Above this threshold, the expected scaling relations were consistently recovered, implying that radiometer noise was no longer the dominant limitation. Therefore, we adopted a one-hour scaled S/N of 1139 (cf. Table~\ref{tab:epochs}) as the threshold for epoch selection and used only those epochs with S/N above this value for the jitter analysis. We did not perform the analysis in Band 5 because the S/N values for those observations were too low to facilitate jitter analysis.

\begin{table}[]
    \centering
    \footnotesize
    
     \caption{Table shows all Band 3 observation used in this work. The first column denotes the uGMRT cycle at which the observation is done, and the second column shows the MJD of the observation. The third column is the one-hour scaled S/N of that epoch. Last column shows the ECORR values (see Section~\ref{subsec:models}) scaled to one hour with error associated with it. Rescaling to one hour is done for the ease of comparison with previous studies. }
    \label{tab:epochs}
    \begin{tabular}{c c c c}
    \hline
    \hline
      Cycle   & MJD & Scaled S/N & ECORR (ns) \\
      \hline
      \hline
        
     \hline \\
        42 & 59746 & 1504 & 89$^{+7}_{-7}$ \\ \\
     \hline \\
        43 & 59954 & 1602 & 97$^{+6}_{-6}$ \\ \\
        43 & 59974 & 1464 & 99$^{+10}_{-6}$ \\ \\
        43 & 59993 & 1152 & 87$^{+8}_{-7}$ \\ \\
        43 & 60005 & 1059 & 92$^{+6}_{-6}$ \\ \\
        43 & 60033 & 1035 & 82$^{+8}_{-8}$\\ \\
     \hline \\
        44 & 60070 & 1516 & 90$^{+7}_{-7}$ \\ \\
        44 & 60089 & 1749 & 94$^{+7}_{-6}$ \\ \\
        44 & 60107 & 1984 & 81$^{+5}_{-5}$ \\ \\
        44 & 60128 & 1949 & 108$^{+7}_{-6}$ \\ \\
     \hline \\
        45 & 60248 & 1329 & 77$^{+8}_{-7}$\\ \\
        45 & 60267 & 1004 & 88$^{+10}_{-10}$\\ \\
        45 & 60288 & 1941 & 111$^{+7}_{-7
        }$ \\ \\
        45 & 60349 & 1099 & 108$^{+9}_{-8}$ \\ \\
        45 & 60371 & 2164 & 91$^{+7}_{-6}$ \\ \\
     \hline \\
        46 & 60444 & 1712 & 96$^{+7}_{-7}$ \\ \\
        46 & 60464 & 2248 & 102$^{+7}_{-6}$ \\ \\
        46 & 60489 & 1596 & 82$^{+6}_{-5}$ \\ \\
        46 & 60505 & 1716 & 89$^{+7}_{-6}$\\ \\
        46 & 60526 & 1332 & 87$^{+7}_{-7}$ \\ \\
        46 & 60574 & 1288 & 83$^{+11}_{-11}$ \\ \\
     \hline \\
        47 & 60650 & 3186 & 97$^{+5}_{-5}$ \\ \\
        47 & 60671 & 1540 & 108$^{+6}_{-6}$ \\ \\
        47 & 60693 & 2028 & 97$^{+6}_{-5}$\\ \\
        47 & 60737 & 1012 & 83$^{+9}_{-9}$ \\ \\
        47 & 60759 & 1139 & 93$^{+8}_{-7}$ \\ \\
        \hline
    \end{tabular}
   
\end{table}

\section{Jitter analysis and results}
\label{sec:analysis}
\subsection{Pulse level analysis} \label{subsec: pulse-jitter}
The effect of pulse jitter is observed in pulsar timing as an excess scatter in the timing residuals that cannot be explained by the formal TOA uncertainties originating from the system noise. This excess arises from epoch-to-epoch variations in the pulse profile shape relative to the fixed template profile. Consequently, the effect of jitter should be detectable not only at the timing level but also directly in the pulse profile domain. Jitter causes a redistribution of the on-pulse energy among neighboring phase bins, leading to increased variance in the on-pulse region. As a result, the on-pulse variance contains contributions from both the radiometer noise and the jitter noise, whereas the off-pulse variance is dominated solely by radiometer noise. In addition, the on-pulse variance is expected to remain approximately constant with increasing bandwidth, since individual pulses are assumed to be correlated over a relatively wide frequency range. 
On the other hand, the variance on the averaged on-pulse bins, summed over T (sub-integration time) is expected to scale as $T^{-1}$, reflecting the averaging of statistically independent pulses.   

\begin{figure}
    \centering
    \includegraphics[width=1
    \linewidth]{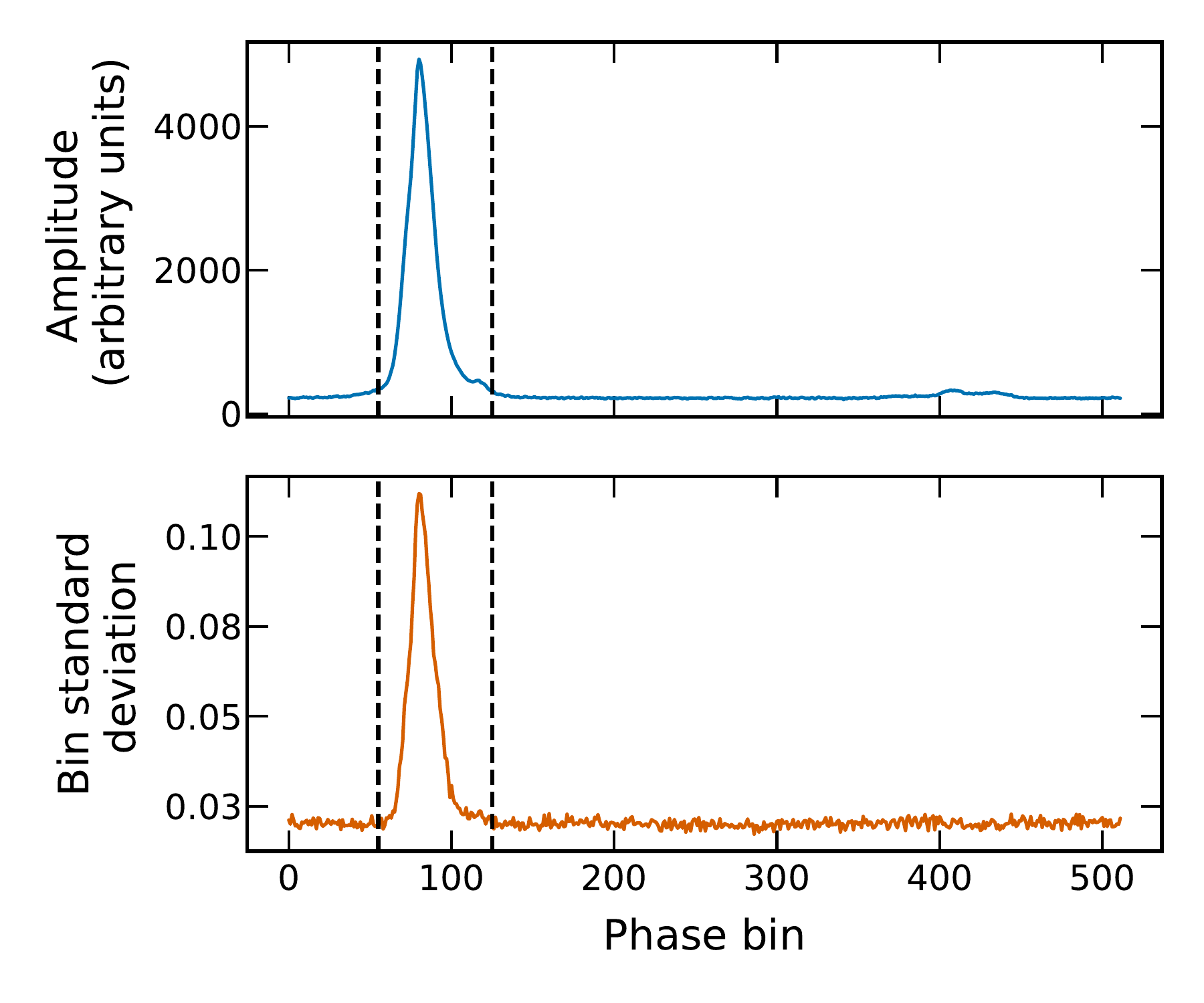}
    \caption{The top panel shows the pulse profile for epoch 60650, and the bottom panel shows the corresponding standard deviation of each phase bin. The on-pulse region is the region between the black dotted lines.}
    \label{fig:fig1}
\end{figure}

\begin{figure}[h]
    \centering
    \includegraphics[width=.95
    \linewidth]{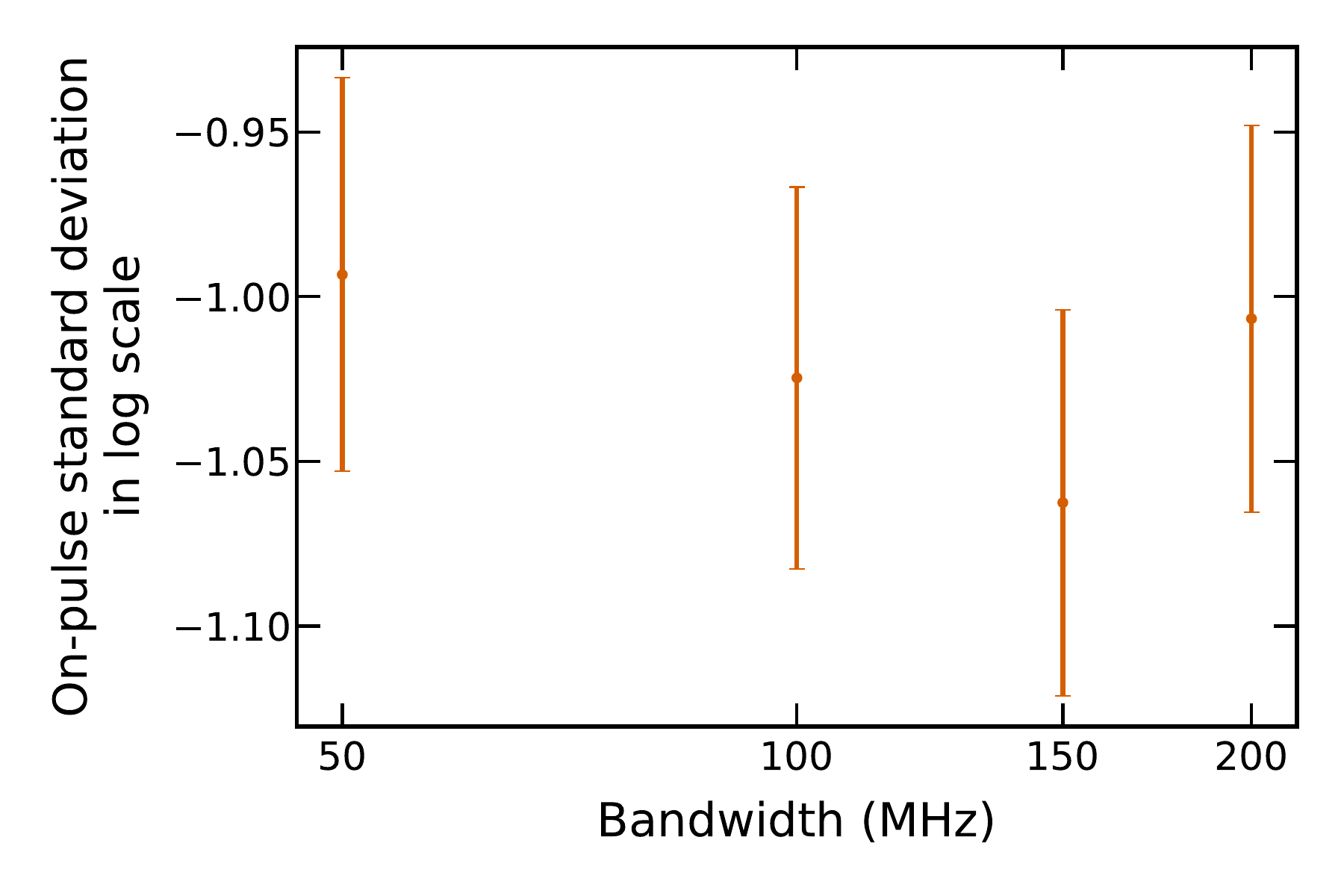}
    \caption{On-pulse standard deviation considering different bandwidths for epoch 60650.}
    \label{fig:fig2}
\end{figure}

\begin{figure}[h]
    \centering
    \includegraphics[width=.95
    \linewidth]{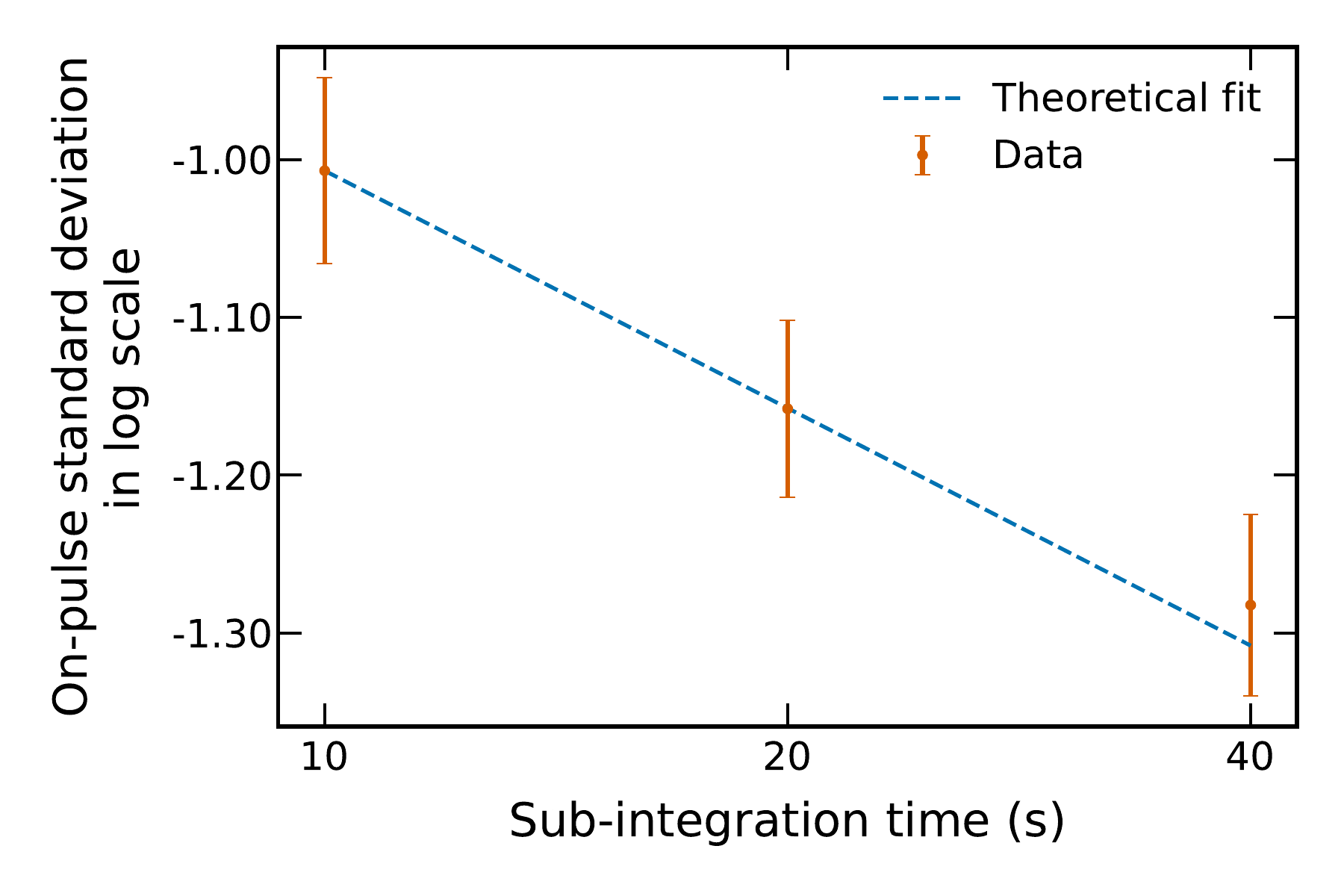}
    \caption{On-pulse standard deviation for different sub-integration times for epoch 60650. The dotted line is the expected scaled-down trend for white noise. }
    
    \label{fig:fig3}
\end{figure}

Therefore, to check for the presence of jitter, the standard deviation of each phase bin is calculated by evaluating the standard deviation of the intensity values in that bin across all sub-integrations. The first panel of Figure~\ref{fig:fig1} shows the pulse profile for epoch 60650, a high-S/N epoch. The second panel shows the standard deviation for each bin in this epoch. The region between the black dotted line is the on-pulse region we are considering (on-pulse lies in the bin range from 55 to 125). We detected that the standard deviation of the on-pulse region is larger than that of the off-pulse region.
To determine how the on-pulse standard deviation scales with sub-integration time and bandwidth,
 we selected 10 bins on either side of the phase bin exhibiting the maximum standard deviation and calculated the mean and the scale parameter (measure of spread of values in selected bins) of the standard deviation values within those bins.
The latter is taken as the uncertainty associated with each point in Figure~ \ref{fig:fig2} and Figure~ \ref{fig:fig3}.
The non-variation of the on-pulse standard deviation with bandwidth points towards the fact that intensity variation is correlated over the observing bandwidth. 
For some epochs, we found an initial decrease of the on-pulse standard deviation, and later it asymptotically reaches a constant level. These results 
strongly indicate the presence of jitter noise for the selected epochs for this pulsar. In the next section, we describe how this jitter is quantified and the noise models used for its characterization.

\subsection{Traditional jitter measurements}
\label{subsec:trad}

\begin{figure}
    \centering
    \includegraphics[width=1
    \linewidth]{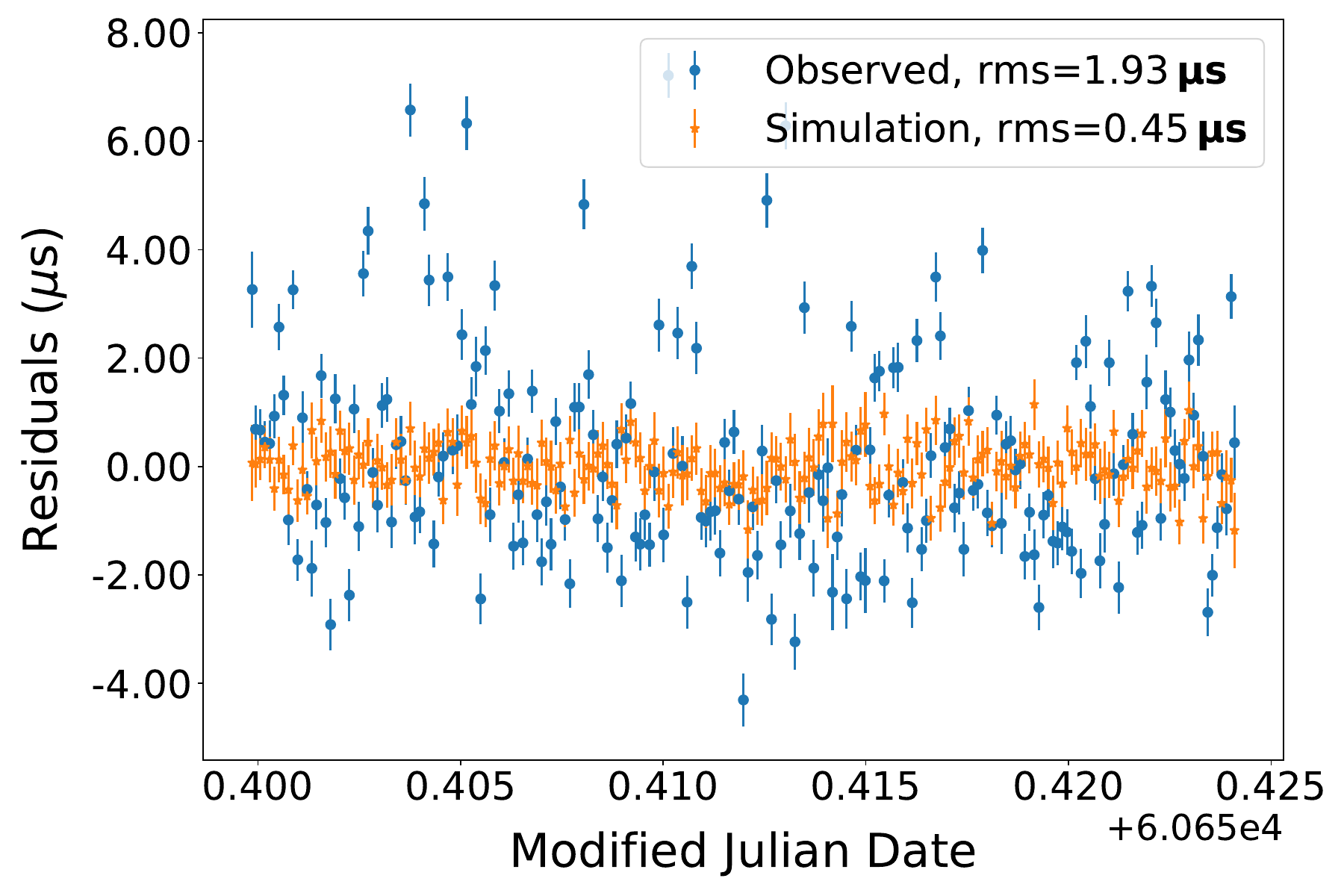}
    \caption{The observed residuals (blue) and the simulated residuals (orange) with \texttt{libstempo} for the observation on the epoch 60650. The RMS values of the observed and the simulated residuals are mentioned in the legend. The quadrature difference between observed and simulated RMS provides a measure of jitter.} 
    \label{fig:fig4}
\end{figure}

 Traditionally, the jitter amplitude is estimated \citep{Parthasarathy2021} by obtaining the quadrature difference between the root-mean-square (rms) of the frequency-averaged timing residuals ($\sigma_{obs}$) and the expected contribution from radiometer noise ($\sigma_{rad}$). 
Based on this {\it ansatz}, the jitter noise is obtained by: 
\begin{equation}
\sigma_J^2(T) = \sigma_{\mathrm{obs}}^2(T) - \sigma_{\mathrm{rad}}^2(T)
\label{eq:eq1}
\end{equation}
where $T$ is the length of sub-integration. $\sigma_{rad}$ can be estimated from the simulated TOAs, which can be obtained using \texttt{toasim} module from the \texttt{libstempo}\footnote{https://github.com/vallis/libstempo} package, with the frequency-averaged TOAs and errors as input.  
\texttt{toasim} module allows the generation of simulated TOAs that are consistent with a given timing model defined by the input parameter file. The \texttt{add\_efac} option can be used to introduce radiometer noise into the simulated timing residuals. The simulation is performed to estimate how much the observed TOA can vary due to radiometer noise alone. From the timing analysis, we obtain the TOA along with its associated uncertainty, where the uncertainty represents the contribution from radiometer noise. In the simulation, we incorporate the stochastic nature of
radiometer noise by generating random noise realizations based on the measured TOA uncertainties and adding them to the observed TOAs. The resulting distribution of simulated TOAs represents the expected variation caused by radiometer noise. This noise is sampled from a Gaussian distribution with TOA errors as the standard deviation. We then obtain $\sigma_{rad}$ from 1000 realizations and then calculate the jitter amplitude using Equation~\ref{eq:eq1}. This method works on the assumption that the rms of simulated TOAs completely takes care of the radiometer noise. 
The observed frequency-averaged timing residuals (represented by blue symbols) and the simulated residuals (represented by orange symbols) for one epoch corresponding to MJD 60650 are shown in Figure~\ref{fig:fig4}. The figure shows that the TOA uncertainties are insufficient to explain the scatter observed in the frequency-averaged residuals, indicating that the excess scatter is likely due to pulse jitter. 
Since the jitter noise behaves like white noise, its amplitude can be scaled using the following relation:
\begin{equation}
    \sigma_J(T_{1})=\sigma_J(T_{2})\sqrt{\frac{T_2}{T_1}}
\end{equation}
Since the jitter amplitude depends on the sub-integration time, we rescale it to an equivalent one-hour integration using the above relation to facilitate comparison with early studies.

\subsection{Models of jitter noise}
\label{subsec:models}

We use Bayesian inference~\citep{Trotta17} to estimate the jitter amplitude and subsequently compare these results with those obtained using the traditional method described earlier in sub-section \ref{subsec:trad}. This sub-section describes the three noise models used for  Bayesian analysis. \par
The observed timing residuals can be modeled as the sum of various deterministic and stochastic noise sources. We assume that our timing residual vector $\delta\textbf{t}$ is free from all deterministic terms and can be completely described by white noise terms. Based on this assumption,  our likelihood function can be written as follows~\citep{Lentati2014}:
\begin{equation}
L(\boldsymbol{\delta t} \mid \theta) =
\frac{1}{\sqrt{(2\pi)^{N_{\mathrm{TOA}}}\,\det\!\left(\mathbf{C}(\theta)\right)}}
\exp\left(
-\frac{1}{2}\,
\boldsymbol{\delta t}^{T}
\mathbf{C}^{-1}(\theta)
\boldsymbol{\delta t}
\right)
\end{equation}
where $N_\mathrm{TOA}= N_\mathrm{subint}\times N_\mathrm{subband}$ and $\textbf{C}(\theta)$ is the covariance matrix with $\theta$ as the set of parameters of the model. \par
In pulsar timing analyses, white noise is commonly characterized using three parameters: EFAC, EQUAD, and ECORR~\citep{Kikunaga2024,Kulkarni2024,Verbiest_IPTA_2016}.
EFAC (error factor) is a multiplicative factor applied to the TOA uncertainties to account for a possible underestimation of the 
radiometer noise. EQUAD (error added in quadrature) represents an additional white-noise term that is added in quadrature to the TOA uncertainties, accounting for excess variance not captured by the TOA errors. This term describes stochastic fluctuations that are uncorrelated between both sub-bands and sub-integrations. ECORR is also added in quadrature; however, it is fully correlated across frequency sub-bands within a given sub-integration and uncorrelated between different sub-integrations. Consequently, the TOAs corresponding to different sub-bands of the same sub-integration share the same ECORR contribution. Since the pulse jitter is expected to be correlated across the observing bandwidth but uncorrelated in time, ECORR naturally represents jitter noise in wideband timing analyses. In contrast, when only a single sub-band is considered, EQUAD serves as an effective proxy for jitter noise. The frequency dependence of the EQUAD parameter can provide insights into the possible frequency dependence of pulse jitter noise. Such frequency-dependent behavior may be related to changes in the magnetospheric emission processes and emission heights sampled at different radio frequencies. We used the following three models to estimate the jitter amplitudes similar to our previous work~\citep{Kikunaga2024}:

\begin{itemize}

\item \textbf{Model I} \\
 Common EFAC and ECORR throughout the band, where the covariance matrix is given by
 \begin{equation}
     \textbf{C}_\mathrm{{ij}}=F^2\sigma^2_{\mathrm{TOA,i}} \delta_\mathrm{{ij}}+J^2\delta_\mathrm{{t_{i}t_{j}}}
     \label{eq:eq4}
 \end{equation}
\item \textbf{Model II} \\
Common EFAC throughout the sub-bands and different EQUAD for each sub-band
\begin{equation}
    \textbf{C}_\mathrm{ij}=(F^2\sigma^2_\mathrm{TOA,i}+Q^2_\mathrm{\nu_\mathrm{i}})\delta_\mathrm{ij}
    \label{eq:eq5}
\end{equation}
\item \textbf{Model III} \\
Different EQUAD for each sub-band without EFAC
\begin{equation}
    \textbf{C}_\mathrm{ij}=(\sigma^2_\mathrm{TOA,i}+Q^2_\mathrm{\nu_i})\delta_\mathrm{ij}
    \label{eq:eq6}
\end{equation}

\end{itemize}
 where, $\sigma_\mathrm{TOA_i}$ is the uncertainty of  $i^{th}$ TOA. The parameter $F$ represents EFAC, and $J$ represents ECORR. $Q_\mathrm{\nu_i}$ denotes the EQUAD corresponding to  $i^{th}$ frequency sub-band and $t_i$ represents the sub-integration of $i^{th}$ TOA. We used a uniform prior distribution in $[0.5,5.0]$ for EFAC and a log-uniform prior distribution in $[-10.0,-5.0]$ for EQUAD and ECORR. Noise models including EFAC, EQUAD, and ECORR were not adopted because the EQUAD parameter could not be constrained by the data.

\subsection{Bayesian analysis}
\begin{figure}
    \centering
    \includegraphics[width=1.09\linewidth]{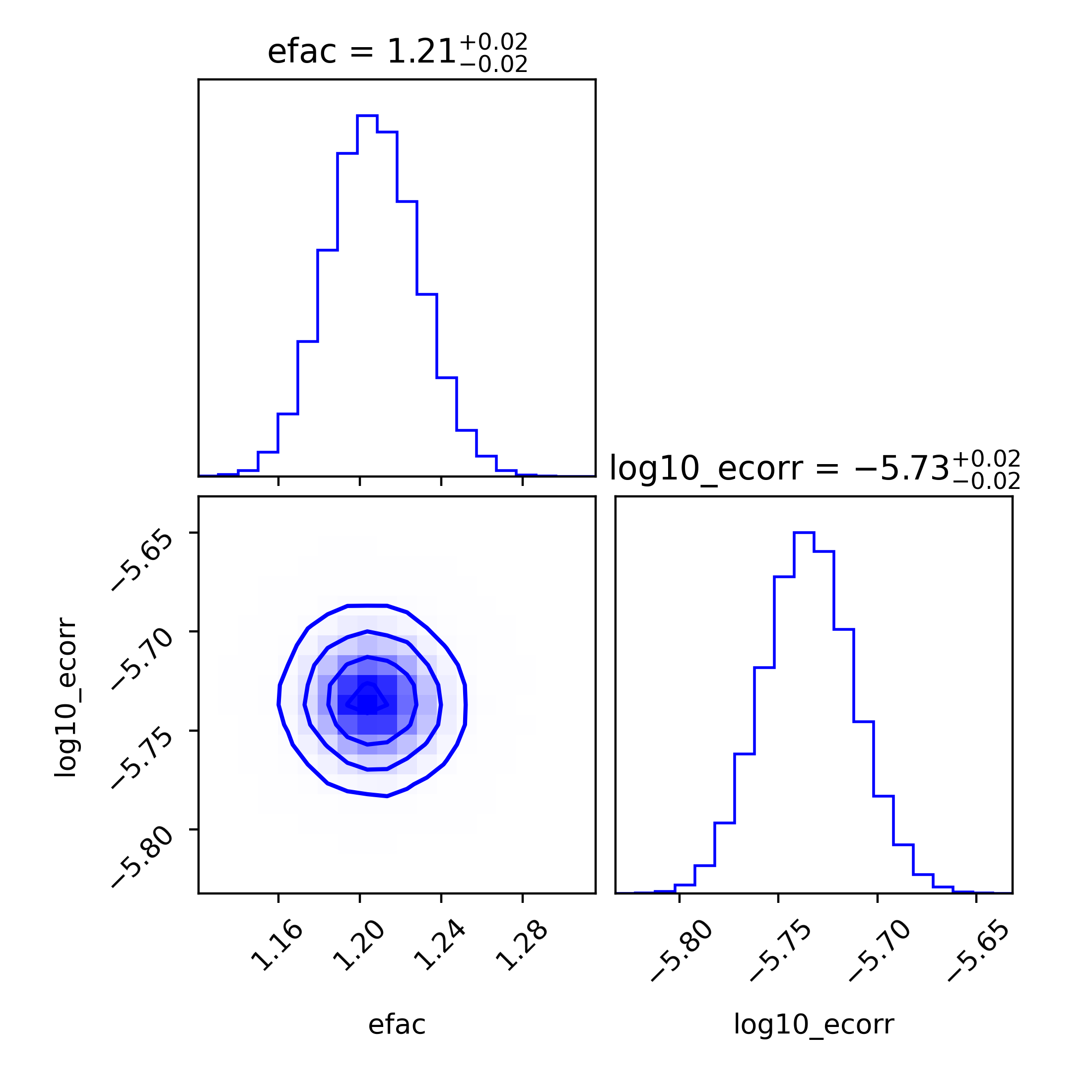}
    \caption{Posterior distribution of the parameters of Model I for epoch 60650 with 60\%, 90\%, 99\% credible intervals. The point estimates shown above represent the median values and marginalized 68\% credible intervals.}
    \label{fig:fig5}
\end{figure}

To estimate the jitter amplitude using Bayesian inference, we employed the \texttt{libstempo} and \texttt{ENTERPRISE} \citep{ellis_2020} Python packages. \texttt{ENTERPRISE} models the covariance between different noise components and estimates their parameters via Bayesian inference, providing the posterior distributions of the parameters for each of the three models considered in this work as elaborated in Section~\ref{subsec:models}.  We used the sub-banded TOAs and the par file modified with epoch -wise-DMs obtained from \texttt{DMCalc} \citep{Kk_2021} as input.
The posterior sampling and parameter estimation for EFAC, EQUAD, and ECORR were carried out using the Markov Chain Monte Carlo (MCMC) sampler implemented in the \texttt{PTMCMCSampler} package \citep{ellis_2017}. Bayesian analyses were performed for all selected epochs. We observed that the inferred noise parameters are poorly constrained when the S/N of the sub-integration or individual sub-bands is low. A representative set of posterior distributions, including the marginalized posteriors of EFAC and ECORR (Model I) for epoch 60650, are shown in Figure~\ref{fig:fig5}.  

\par
\begin{figure}[h]
    \centering
    \includegraphics[width=1.05\linewidth]{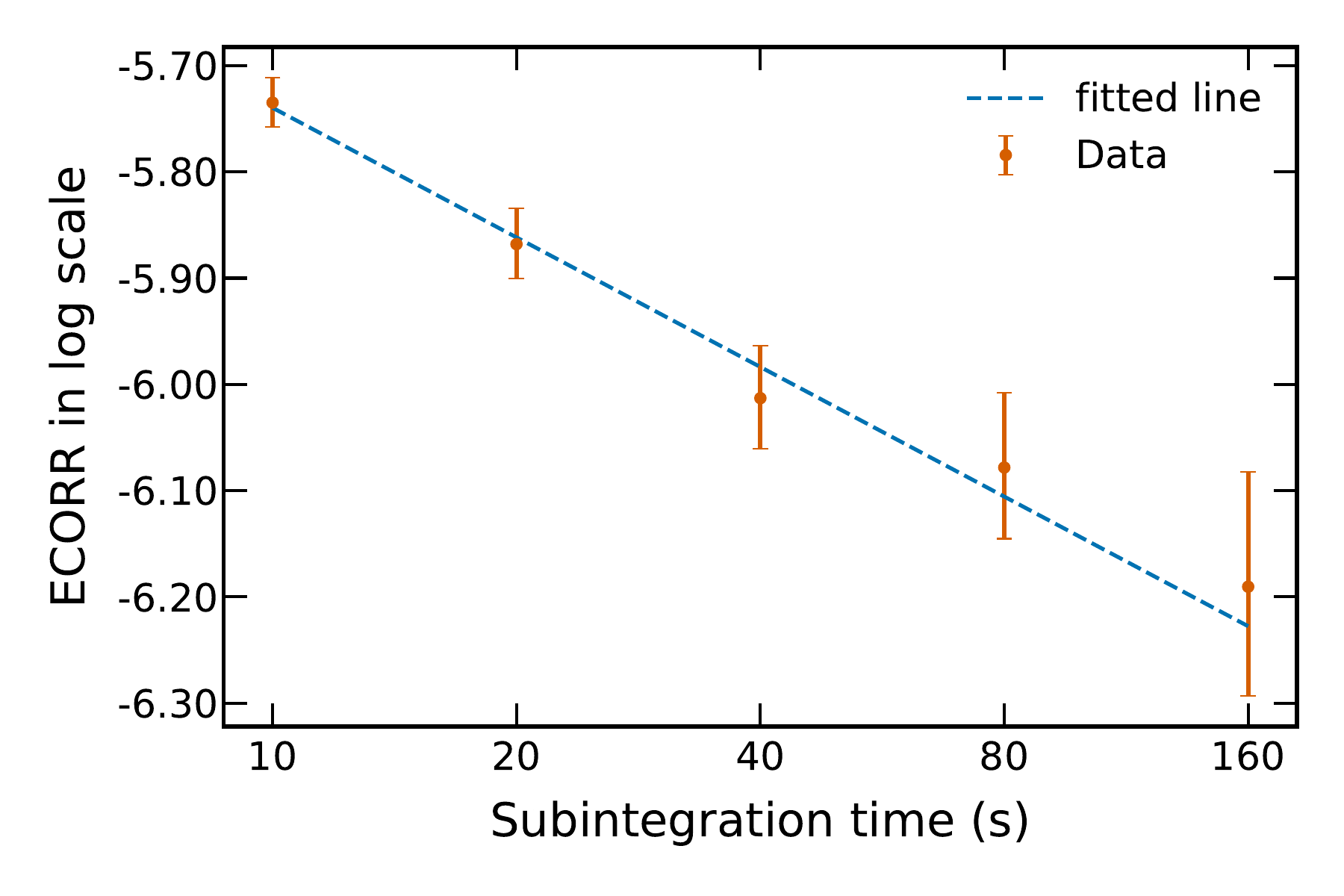}
    \caption{ECORR values for the epoch 60650 from model I are plotted against different subintegration times. The X-axis is on a log scale. The blue dotted line is the 
    fit, for which the fitted function is proportional to the inverse square root of sub-integration time.  Slope of fit is $-0.45^{+0.05}_{-0.05}$.}
    \label{fig:fig6}
\end{figure}
\begin{figure}[h]
    \centering
    \includegraphics[width=1.05\linewidth]{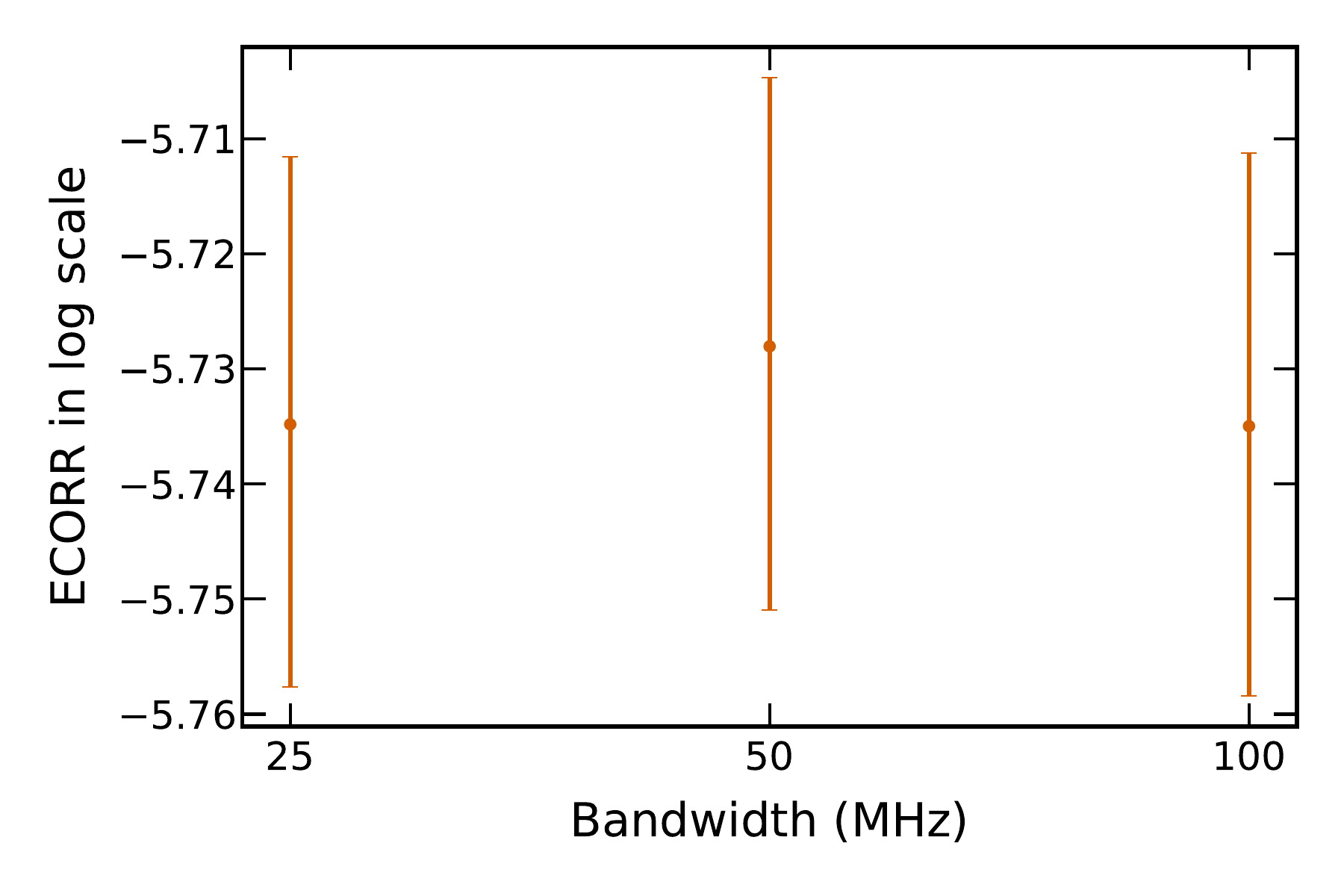}
    \caption{ECORR values for the epoch 60650 from the Model I plotted against different bandwidths. The X-axis is on a log scale. ECORR values are for 10-second sub-integration times for each considered bandwidth }
    \label{fig:fig7}
\end{figure}

To test whether the ECORR values follow the expected white noise scaling with sub-integration time and exhibit correlated behavior across the observing bandwidth, we performed additional analyses using different data configurations. Specifically, ECORR was obtained using different sub-integration lengths, and by frequency averaging the data from the initial eight sub-bands to four and two sub-bands. Figure~\ref{fig:fig6} shows the scaling of ECORR with sub-integration time, which is consistent with the white noise trend that we are expecting. The blue dotted line represents the fit to this trend, which is theoretically expected to have a slope of −0.5. In our case, we obtain a slope of $-0.45^{+0.05}_{-0.05}$. 
Figure \ref{fig:fig7} shows the constant nature of ECORR over the bandwidth, which is consistent with the correlated nature of jitter over the bandwidth. Assuming that ECORR follows the expected scaling with higher sub-integration time, we rescale all measured values to a reference integration time of one hour. 

\par The ECORR estimates for the different suitable epochs are presented in the last column of Table~\ref{tab:epochs}. The uncertainties given are at the $1\sigma$ level. It should be noted that this is the first estimate of jitter for \src at this frequency band. Taking into account all the ECORR values given in this table, the mean ECORR value of \src at 400 MHz turns out to be $93\pm9$ ns.
 
\par
\begin{figure}[h]
    \centering
    \includegraphics[width=1.05\linewidth]{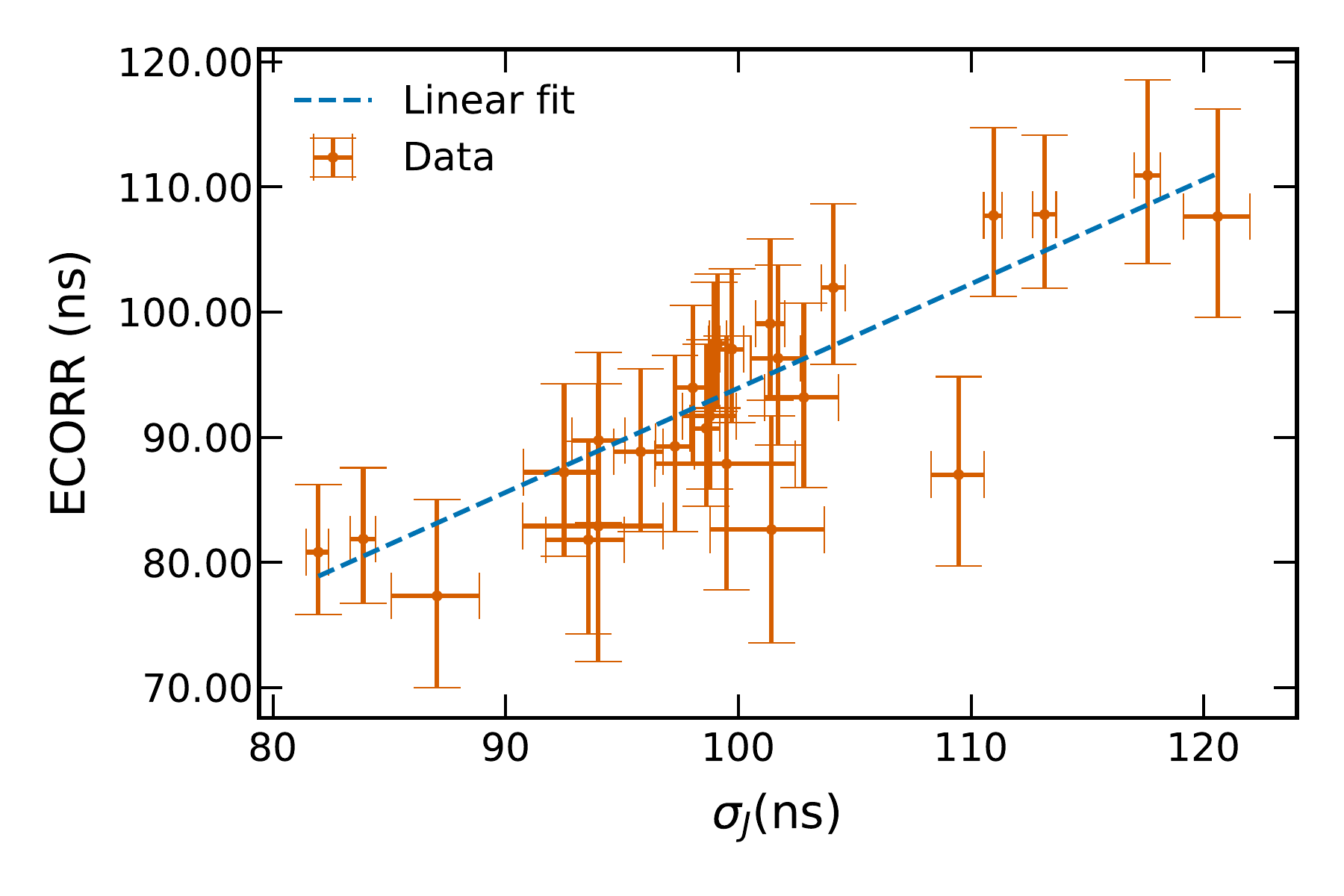}
    \caption{Comparison of the jitter amplitudes obtained from the Bayesian analysis (ECORR) and from the traditional method. For quantitative comparison, a linear fit was performed to examine the correlation between the two, as shown by the blue dashed line, and the slope of the fit was obtained to be $0.83^{+0.14}_{-0.14}$. The traditional method does not account for correlation or scaling with EFAC. This explains the deviation of slope from the ideal 1 value.}
    \label{fig:fig8}

\end{figure}

For consistency check, we have plotted in Figure \ref{fig:fig8} the ECORR values obtained from Model I with the jitter amplitude from the traditional way explained in section \ref{subsec:trad}. The blue dotted line shows the linear fit with slope $0.83 \pm 0.14$. The slight offset from the ideal slope of unity can be understood in terms of the different treatment of excess noise in the two methods. In the traditional approach, additional scatter in the measured ToAs is  attributed to jitter, whereas the Bayesian analysis provides an additional constraint through the ECORR term, which accounts for correlated variations. Consequently, the traditional method can yield a larger inferred jitter amplitude than the Bayesian method, potentially resulting in a slope below unity when the two estimates are compared.

. 

\subsection{Frequency dependence of jitter amplitude}
To investigate the frequency dependence of the jitter amplitude, we estimated the EQUAD values using Model II and Model III. In these models, EQUAD represents an additional white-noise term added in quadrature to the TOAs of each sub-band, which we interpret as the jitter amplitude, consistent with the definition adopted in the traditional method described in Section \ref{subsec:trad}. \par
From our Bayesian analysis, we observed that the EQUAD values obtained from Model II are not well constrained. Consequently, their large uncertainties prevent us from drawing meaningful conclusions about the frequency dependence of the jitter amplitude in this case. In contrast, Model III yields well-constrained EQUAD estimates, allowing a more reliable investigation of its frequency dependence. 
It is evident that S/Ns have a considerable impact on the jitter, and since we have considered only those sub-bands which have S/N greater than 200. 

\begin{figure}
    \centering
    \includegraphics[width=1.05\linewidth]{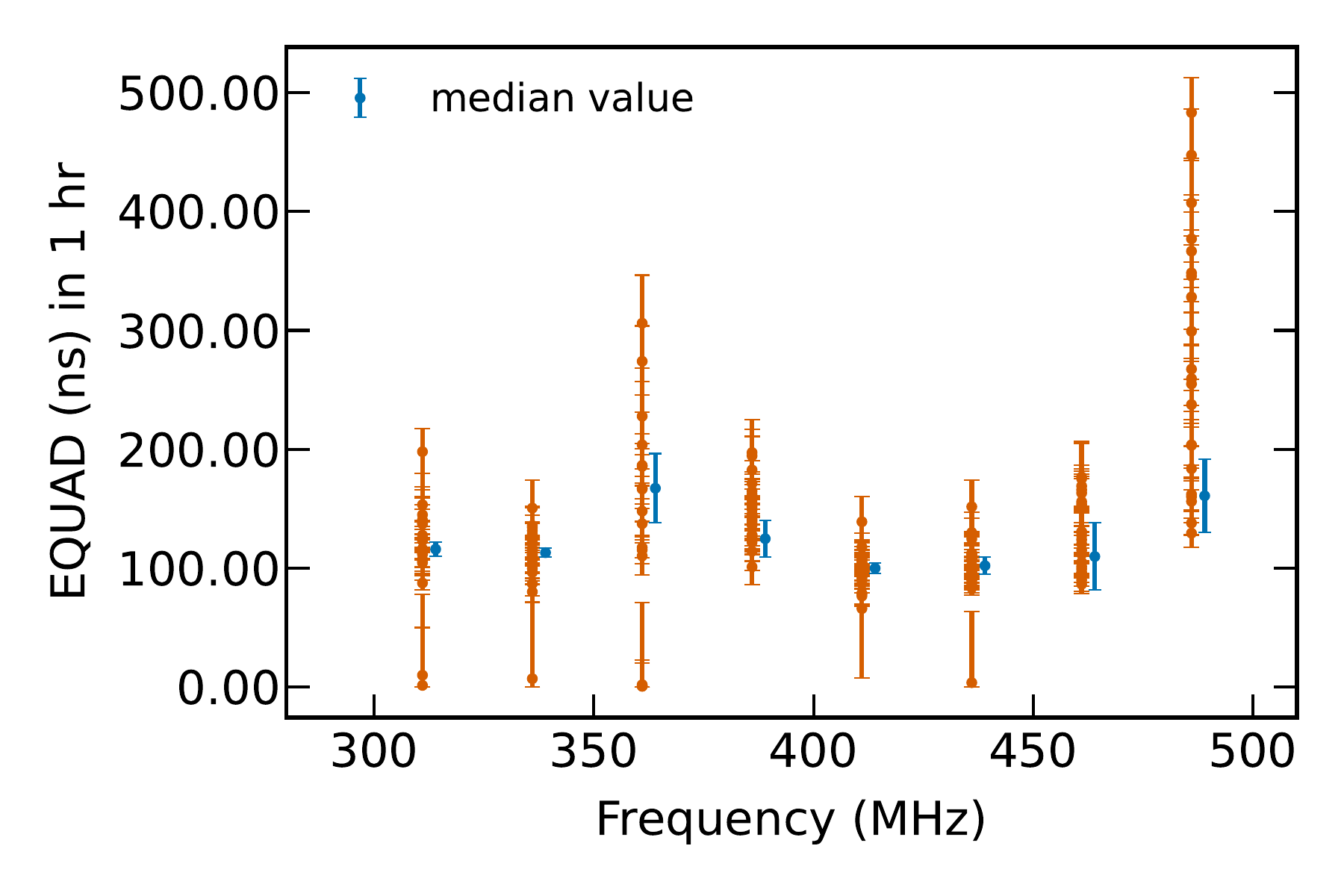}
    \caption{Plot of EQUAD values scaled to 1 hour from every epoch using Model III plotted as a function of frequency. Blue lines are the median EQUAD values which is plotted with 3 MHz offset to the right. }
    \label{fig:fig9}
\end{figure}

\begin{figure}
    \centering
    \includegraphics[width=1.05
    \linewidth]{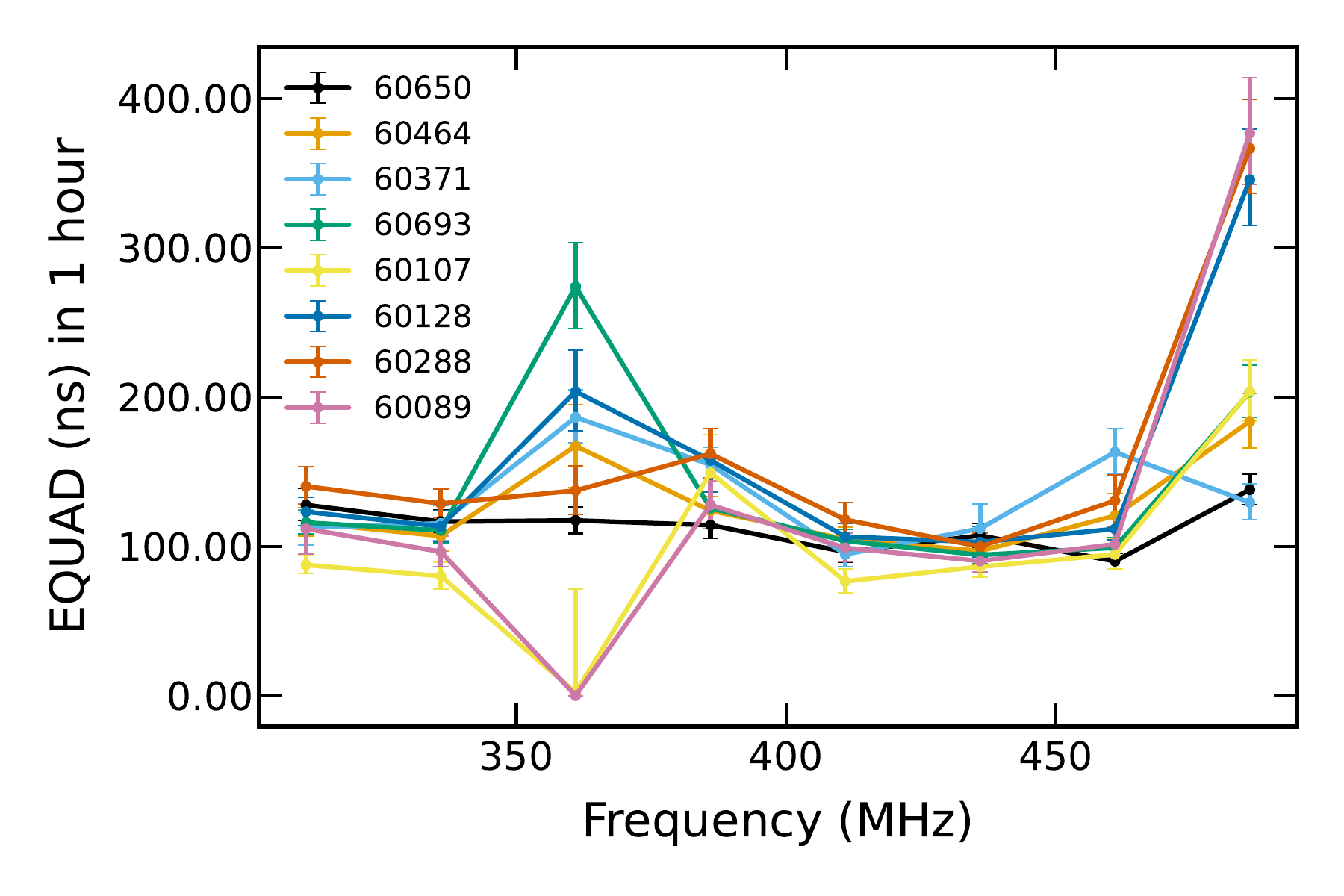}
    \caption{Frequency evolution of EQUAD values scaled to 1 hour for different epochs depicted using the different colors.}
    \label{fig:fig11}
\end{figure}

The EQUAD values for the eight sub-bands were obtained for all epochs with sufficient S/N and plotted as a function of frequency in  Figure \ref{fig:fig9}. No significant systematic trend is evident from a visual inspection. For visualizing the absence of the trend more robustly, we computed the median of EQUAD values in each frequency band and used the dispersion of the values as the associated uncertainty for each subband, which are the blue lines in Figure \ref{fig:fig9}. 
This further reaffirms the absence of frequency dependence of EQUAD in the uGMRT frequency range. 
To examine whether the frequency dependence exhibits a time variation, we also investigated the evolution of EQUAD dependence on frequency over different selected time epochs.
Figure \ref{fig:fig11} shows the frequency evolution of EQUAD values of the first eight highest S/N epochs, with the different colors representing the various epochs. 
The highest S/N epochs are chosen so that the constrained EQUAD values can be estimated over different subbands across different time epochs. It is clear from the figure that the EQUAD values do not follow a clear trend with frequency or a systematic evolution with time. Although an apparent increase is seen toward the 361 MHz frequency band, this is likely because that sub-band has a low S/N for most epochs. The low S/N may be attributed to RFI contamination, as the off-pulse standard deviation is found to be higher in this sub-band than in the other sub-bands. Similarly, a large variability in the EQUAD values is observed around 486 MHz. This variability could be due to RFI contamination or a bandpass effect, as the telescope sensitivity generally decreases toward the edges of the observing bandwidth.

\subsection{Correlation study on sub-banded ToA residuals}
\begin{figure}
    \centering
    \includegraphics[width=1\linewidth]{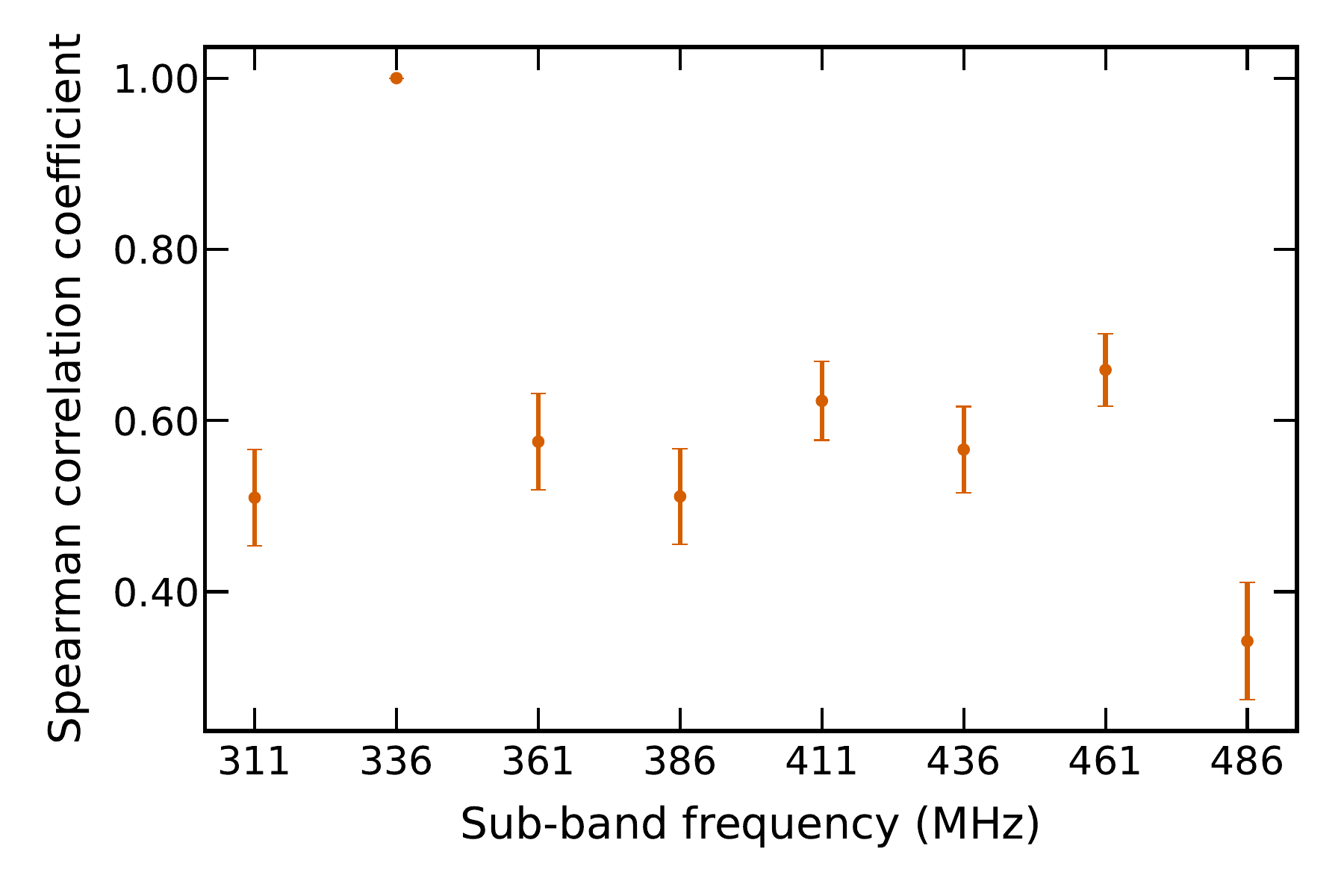}
    \caption{Spearman correlation between the residuals of sub-bands of the epoch 60650. The correlation for each sub-band is computed relative to the sub-band with a central frequency of 336 MHz. }
    \label{fig:fig12}
\end{figure}

As the jitter noise is assumed to be correlated across the observing bandwidth, significant correlations are expected between TOA residuals obtained from different sub-bands, especially those that are adjacent or closely spaced.  Figure~\ref{fig:fig12} shows the Spearman correlation coefficients for eight sub-bands (each 25 MHz wide) at the epoch MJD 60650, with the second sub-band used as the reference. The uncertainties are estimated by bootstrapping the 10~s sub-integration ToA residuals of each sub-band. Contrary to our expectations from the correlated nature of jitter, and early study in other pulsars \citep{Parthasarathy2021}, we obtained Spearman correlation coefficients of about 0.6 suggesting only partial correlation.\par 

A possible explanation for this result would be that the noise in each sub-band may consist of both a correlated component (jitter) and an additional uncorrelated component~\citep{Oslowski11, Kulkarni2024}. In other words, each sub-band could contain a common noise term shared across the observing bandwidth, along with a sub-band-specific noise contribution. The common component would manifest as ECORR in the noise model, while the uncorrelated component would reduce the measured Spearman correlation coefficient. We assume that the amplitude of the correlated noise component is larger than that of the uncorrelated noise. Consequently, the Bayesian analysis yields ECORR values consistent with expectations (Inverse-square-root scaling with sub-integration time, and its independence from bandwidth), but the correlation studies yield Spearman correlation coefficients with partial correlation as obtained earlier. \par

\subsection{Jitter and the DM variations}
\label{sec:dm}

\citet  {Parthasarathy2021} had raised the concern of jitter being the limiting factor in the DM precision measurements of MSPs from their short-duration observations, based on their analysis in the high-frequency regime. We had investigated this on PSRJ0437$-$4715, and found that for high S/N, short duration observations, jitter can affect DM precision which can be overcome by observing for long enough duration optimised for a particular pulsar. ~\citep{Kikunaga2024}. Here we revisit the same question and investigate the effect of jitter on the DM precision for PSR J1744$-$1134 in the low frequency regime (300$-$500 MHz) by obtaining the DM uncertainty for its highest S/N epoch in our dataset. We measured the DM for this single epoch (of 35 min duration) to be
$3.138984~\mathrm{cm^{-3}\,pc}$ with a DM error of $3\times10^{-5}$ units.
Using the jitter estimates given in Table~\ref{tab:epochs}, we calculate the
jitter in a 35 min observation to be about $1.3\times10^{-7}$ s. To estimate the DM precision that jitter alone can limit, we adopt the jitter as the TOA uncertainty and propagate it through the dispersion delay between the band edges. This gives a jitter-induced DM error of approximately $7\times10^{-6}$ units. Jitter therefore accounts for only a small fraction
of the DM uncertainty, and we conclude that it is not the limiting factor for the DM precision of PSR J1744$-$1134 for this integration time.

\section{Discussion and Conclusions}
\label{sec:disc}
We have conducted the first study of the jitter noise in PSR J1744$-$1134 in the frequency range of 300$-$500 MHz using uGMRT observations as part of the InPTA experiments. The results obtained from all three methods, viz., the pulse-level analysis, the traditional method, and the Bayesian analysis described in Section \ref{sec:analysis}, clearly indicate the presence of jitter noise in this pulsar for the suitable S/N epochs.  Given the high S/N of this pulsar at these frequencies, this is a strong arguement for the detailed jitter characterization, which will become critical for high-sensitivity timing studies. \textbf{}\par

Table~\ref{tab:jitter} presents the comparison of the jitter estimates in this work with the previous jitter estimates for \src at various frequencies. Previous studies of jitter in PSR J1744$-$1134 have reported amplitudes that are smaller than the values obtained in this work; however, those measurements were performed primarily at much higher frequency ranges. Notably, our results highlight the importance of jitter characterization for pulsar timing studies at low frequencies. Although we do not yet have substantial evidence, the observed difference in amplitude may hint at the underlying frequency dependence of jitter and the possible presence of a turnover frequency, similar to what was proposed for PSR J0437$-$4715 \citep{Kikunaga2024}. Previously, such frequency-dependent characteristics of jitter have been reported for a large sample of pulsars \citep{Lam2019}. For PSR J1744$-$1134, \citet{wang2024} reported a decrease followed by an increase of the jitter amplitude for the frequency range 820$-$3100 MHz. However, within the uGMRT bandwidth, we did not observe any significant frequency dependence of the jitter amplitude for PSR J1744$-$1134. It may be possible that the frequency dependence of the jitter amplitudes is discernible over wider frequency scales. It should be noted that the weak frequency dependence of jitter in this frequency band can also be attributed to the lesser modulations introduced at these lower frequencies by shorter propagation path in the magnetosphere (considering that lower frequencies are originating from higher emission heights and Radius-to-frequency mapping (RFM) is valid for this pulsar) \citep{Kikunaga2024}.
Additionally, S/N can also be a critical impacting factor in unveiling the frequency dependence of jitter. Interestingly, at 944 MHz, PSR J1744$-$1134 exhibits a much higher modulation index (RMS fluctuation of pulse intensity, normalized by the mean pulse intensity) compared to  PSR J0437$-$4715 for similar DM values \citep{Gitika_2023}. More sensitive simultaneous multi-frequency observations are required for addressing this issue.

\begin{table}[]
    \centering
    \begin{tabular}{ccc}
    \hline\hline
    Paper & Frequency & ECORR (ns) \\
    \hline
    Lam et al. (2019)           & 1500 MHz & $46.5^{+1.3}_{-1.3}$ \\ \\
    Lam et al. (2019)           & 820  MHz & $41.72^{+2.6}_{-2.6}$ \\ \\
    Shannon et al. (2014)       & 1400 MHz & $37.8^{+0.8}_{-0.8}$ \\ \\
    Parthasarathy et al. (2021) & 1284 MHz & $30^{+6}_{-6}$ \\ \\
    Wang et al. (2024)          & 1250 MHz & $29.4^{+0.2}_{-0.2}$ \\ \\
    {\bf This work}             & 400 MHz  & $93^{+9}_{-9}$ \\ \\
    \hline
    \end{tabular}
    \caption{Jitter amplitude (ECORR) scaled to 1 hour reported in different studies at various observing frequencies.}
    \label{tab:jitter}
\end{table}

\citet{Oslowski11} reported that the timing residuals for PSR J0437$-$4715 of two frequency channels were highly correlated for an observation duration of one hour, which was attributed to the presence of jitter. For the same pulsar, \citet{Parthasarathy2021} compared the jitter amplitudes in different frequency subbands, and observed a correlation between adjacent frequency bands which gradually exhibited a decorrelation as the separation between the channels increased. 
\citet{Kulkarni2024} also investigated the correlated jitter behavior of PSR J0437$-$4715 across frequencies and reported a decorrelation of jitter noise with frequency that can be modeled by a power law. They apply this jitter decorrelation relation to explain variations in chromatic noise. Moreover, the similar behavior of the pulse energy decorrelations shows how energy of each sub-bands are correlated, and the jitter parameter decorrelations further emphasizes a strong connection between the two variables.
In this work, we did not detect a significant correlation of the TOA residuals between adjacent frequency subbands for PSR J1744$-$1134.
This low correlation can be due to the presence of uncorrelated noise in each sub-bands along with the correlated jitter noise. Similar behavior was also discussed by \citet{Oslowski11} by implementing the principal component analysis technique. Additionally, the observed decorrelation may have a physical origin in the pulsar magnetosphere. In the 300$-$500 MHz frequency range, emission is expected to originate from a range of altitudes due to radius-to-frequency mapping, with lower frequencies arising from higher emission heights~\citep{Komesaroff1970, Cordes1978}. Variations in plasma conditions across these regions can lead to imperfect coherence of the emission process, resulting in partial decorrelation of jitter across sub-bands. An ECORR+EQUAD model provides insight into the correlated and sub-band-specific uncorrelated noise components. However, a larger number of independent sub-integrations is required to reliably constrain these two parameters.

Furthermore, \citet{Lam2019} investigated the time dependence of the jitter behavior to confirm the stability of the jitter noise. They reported that the frequency dependency of jitter remained stable over time for the case of pulsars like PSR J1713$+$0747, and any deviation can be explained by scintillation effects. This scenario appears to be consistent with our time evolution results for PSR J1744$-$1134. As shown in Figure~\ref{fig:fig11}, even considering the high S/N epochs, the frequency-dependent jitter (EQUAD) did not show significant variation with time. 

In Section \ref{sec:dm}, we investigated the effect of jitter on DM precision measurements. We found that jitter does not affect the DM precision achieved in our high S/N observation of 35 min duration. Though it remains to be seen whether jitter would affect significantly smaller duration observations.

In this work, we conducted the first detailed study of jitter characteristics for PSR J1744$-$1134 - a pulsar favored by most PTAs - at low frequencies. Next-generation telescopes like the Square Kilometer Array (SKA) will reach remarkable TOA precision over short integrations owing to its unprecedented sensitivity. Assuming the systematic timing model parameters to be correctly estimated at this level, the jitter noise is expected to become a dominant contributor to the timing error budget and will limit the achievable timing precision \citep{Oslowski11, Liu2011}. This presents a challenge for short-term timing strategies in next-generation telescopes, where the observation time allocated to each pulsar must be guided not only by the desired S/N, but also by the contribution of jitter noise. This underscores the importance of comprehensive jitter characterization for the pulsars, especially the bright sources, used in pulsar timing experiments.





\section{Acknowledgments}
We acknowledge the support of GMRT staff in resolving technical difficulties and GMRT telescope operators for the observations. CD acknowledges the support of the Department of Space, Government of India. KT is partially supported by JSPS KAKENHI grant Nos. 24H01813, 25K21670, 26H00838 and 26K21724. BCJ acknowledges the support from 
Raja Ramanna Chair fellowship of the Department of Atomic 
Energy, Government of India (RRC $-$ Track I Grant 3/3401 
Atomic Energy Research 00 004 Research and Development 
27 02 31 1002//2/2023/RRC/R\&D-II/13886 and 
1002/2/2023/RRC/R\&D-II/14369). BCJ also acknowledges 
support from the Department of Atomic Energy Government 
of India, under project number 12-R\&D-TFR-5.02-0700. KR is supported by Supported by UGC NET-JRF fellowship.  ZZ is supported by the Prime Minister’s Research Fellows (PMRF) scheme, Ref. No. TF/PMRF-22-7307. NDB is supported by Department of Science and Technology, Government of India vide the DST-WISE fellowship (DST/WISE-PDF/PM-17/2024). SD is supported by SERB MTR/2023/000384.

\bibliographystyle{elsarticle-harv}
\bibliography{references}
\label{lastpage}

\end{document}